\documentclass[aps,prl,twocolumn,showpacs,floatfix,amsmath,amssymb]{revtex4}

\usepackage{graphicx}
\usepackage{dcolumn}
\usepackage{bm}
\usepackage[below]{placeins}
\usepackage{color}
\usepackage{ulem}

\begin{document}
\title{Magnetic spectra in the tridiminished-icosahedron \{Fe$_9$\} nano-cluster by inelastic neutron scattering}
\author{David Vaknin$^{1}$ and Franz Demmel$^2$}
\affiliation{$^1$Ames Laboratory and Department of Physics and Astronomy, Iowa State University, Ames, Iowa 50011, USA\\
$^2$Rutherford Appleton Laboratory, ISIS Pulsed Neutron Facility, Chilton, Didcot, Oxon OX110QX, UK}

\begin{abstract}
Inelastic neutron scattering (INS) experiments under applied magnetic field at low temperatures show detailed low lying magnetic excitations in the so called tridiminshed iron icosahedron magnetic molecule.  The magnetic molecule consists of nine iron Fe$^{3+}$ ($s = 5/2$) and three phosphorous atoms that are  situated on the twelve vertices of a nearly perfect icosahedron.  The three phosphorous atoms form a plane that separates the iron cluster into two weakly coupled three- and six-ion fragments, \{Fe$_3$\} and \{Fe$_6$\}, respectively.  The magnetic field INS results exhibit an $S=1/2$ ground state expected from a perfect equilateral triangle of the \{Fe$_3$\}  triad with a powder averaged $g$-value $=2.00$.  Two sets of triplet excitations whose temperature and magnetic field dependence indicate an  $S=0$ ground state with two non-degenerate $S=1$ states are attributed to the \{Fe$_6$\} fragment. The splitting may result from a finite coupling between the two fragments, single-ion anisotropy, antisymmetric exchange couplings, or from magnetic frustration of its triangular building blocks.  

\end{abstract}
 \email{vaknin@ameslab.gov}
\maketitle

\begin{figure}
\includegraphics[width=3. in]{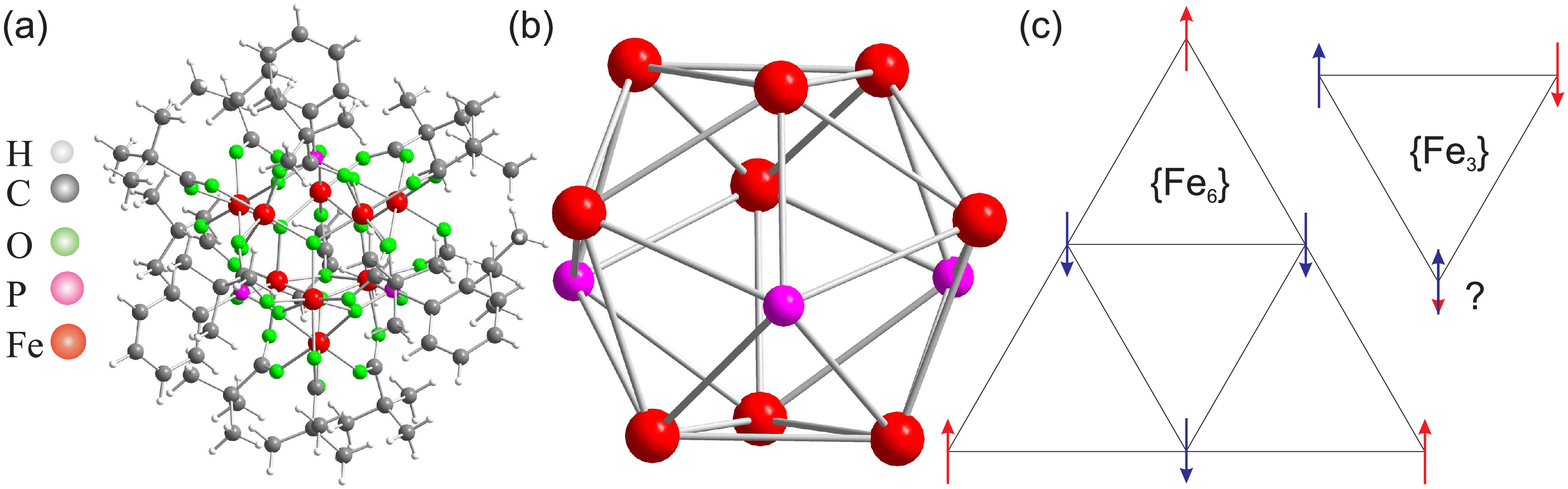}
\caption{(Color online) (a) \{Fe$_9$\} molecule used in this study.  (b) The core icosahedron is occupied by nine iron and three phosphorus ions.  The three phosphorus atoms separate the iron atoms into two \{Fe$_6$\} and \{Fe$_3$\}  fragments. (c) The topology (and exchange pathways) in the  \{Fe$_6$\} and \{Fe$_3$\} clusters. Arrows, representing spins, illustrate geometrical frustration but do not correspond to actual spin orientations in each fragment. }
\label{fig:molecule}
\end{figure}

Single-molecule magnets have been emerging as the necessary ingredients to transform molecular electronics into molecular spintronics, albeit not without challenges in synthesis, characterization, and theoretical modeling\cite{Bogani2008}. Progress in synthesizing, by design, bulk magnetic molecules has brought about new insights into magnetic phenomena in the nanoscale regime \cite{Gatteschi2006,Winepenny2012,Furrer2013}.  A current challenge in this research area is the synthesis of magnetic clusters that can operate as nano-switches. One such hypothetical system is an iron based icosahedron magnetic cluster \{Fe$_{12}$\}, that theoretical calculations show undergoes a first order metamagnetic transition with an associated hysteretic behavior that renders it magnetic nano-switch like properties\cite{Ioannis2005,Schroder2005,Schroder2005b}.  Although the perfect icosahedron has not yet materialized, a closely related molecule of tridimished icosahedron \{Fe$_9$\}, with the chemical formula [(Fe$_9$-$\mu_3$-O)$_4$(O$_3$PPh)$_3$(O$_2$CCMe$_3$)$_{13}$], has been synthesized and characterized\cite{Tolis2006}.  In each \{Fe$_9$\} cluster, Fe$^{3+}$ ions (spins $s = 5/2$) occupy nine of the twelve sites of an icosahedron (see Fig.\ \ref{fig:molecule}) whereas the remaining three vertices are occupied by P atoms of tri-phenylphosphonate. Analysis of exchange paths and susceptibility data indicate that the icosahedron can be divided into two nearly decoupled clusters:  an \{Fe$_6$\} cluster, with an $S=0$ ground state, and an \{Fe$_3$\} triangle with $S=1/2$ ground state. 

Here, we report on the low-lying magnetic excitations of a partially deuterated polycrystalline sample of the \{Fe$_9$\} molecule and their dependence on temperature and applied magnetic field  obtained from inelastic neutron scattering (INS) measurements.  INS  has become a pivotal tool in determining the magnetic spectra of magnetic molecules\cite{Furrer2013} in particular, since the seminal study by Caciuffo and coworker of the \{Fe$_8$\} molecule with frustrated AFM couplings that render the molecule $S =10$ ground state\cite{Caciuffo1998}. Recently, INS has been employed to the Cr$_8$ AFM ring to detect propagation of quantum fluctuations along the ring\cite{Baker2012}, and in the related  \{Cr$_7$Ni\} ring to demonstrate that under external magnetic field an avoided crossing between energy levels of different total-spin quantum numbers can be identified\cite{Carretta2007}. In a systematic study of various magnetic molecules, frustration effects have been classified by use of INS experiments \cite{Baker2012b}.  Employing INS, we monitor the response of the  \{Fe$_9$\} nano cluster  to magnetic field and temperature to obtain a comprehensive picture of the low-lying excitations. 

The INS measurements were performed on approximately 4 g of polycrystalline sample that was sealed under helium atmosphere in a copper can and placed in a dilution refrigerator  with the capability to cool the sample to base temperature of 0.1 K, and to apply magnetic field up to 5 Tesla.   The inelastic neutron spectra were collected on the OSIRIS time-of-flight spectrometer at ISIS, which was set up at a fixed final neutron energy of $E_f = 1.845$ meV that was selected by a pyrolytic graphite PG(002) analyzer\cite{OSIRIS}. The energy resolution, that OSIRIS provides, is approximately 23 to 27 $\mu$eV over the energy-transfer range from zero to 2 meV.\cite{OSIRIS} 
The obtained energy $versus$ momentum-transfer ($E-Q$) slices of the spectra  exhibit distinct excitations that are dispersionless as expected from non-interacting (i.e., zero-dimensional) magnetic molecules. We also note that the variation of the magnetic form factor of Fe$^{3+}$ is less than 5\% over the measured $Q$-range. We therefore integrate intensities over the measured $Q$ range to improve signal statistics, without loss of information.   The theory of INS for magnetic molecules has been extensively detailed in the literature\cite{Waldmann2003,Furrer2013}.

\begin{figure}
\includegraphics[width=2.1 in]{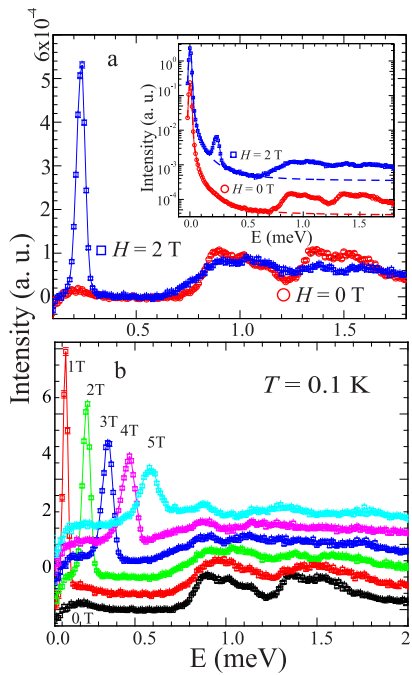}
\caption{(Color online) (a) inset.  Intensity {\it versus} neutron energy loss integrated over the $Q=0.7$ to 1.8 \AA$^{-1}$ for \{Fe$_9$\} at $T = 0.1$ K at 0 and at 2 Tesla.  Dashed line is the instrumental resolution function (including incoherent scattering from our sample). The subtraction of the resolution function from the raw data (inset) yields the intrinsic magnetic spectra. (b) Same as (a)  at  $T=0.1$ K and at various applied magnetic fields showing the emergence of an $S=1/2$ state from the \{Fe$_3$\} cluster, and its effect on the two excited $S=1$ states of the \{Fe$_6$\} cluster.    Spectra at each field are shifted vertically for clarity.}
\label{fig:Data-Reduction}
\end{figure}
\begin{figure}\includegraphics[width=2.1 in]{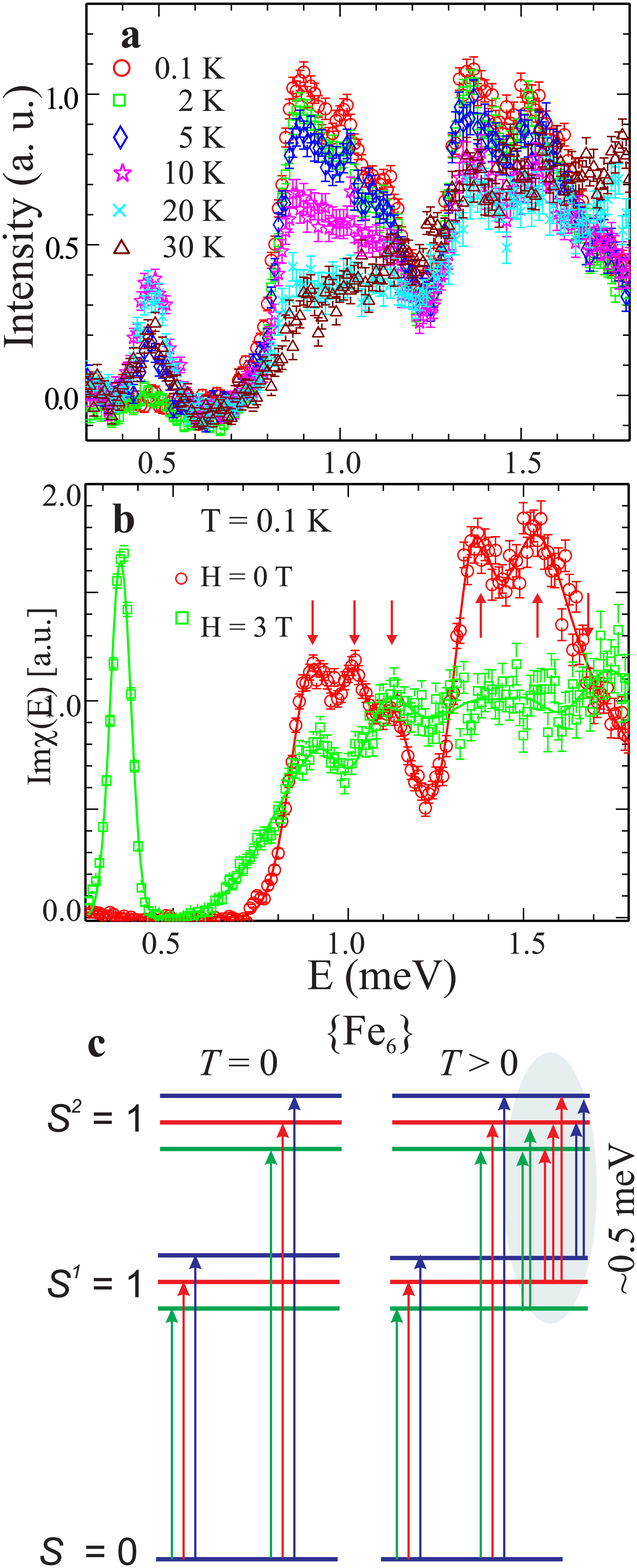}
\caption{(Color online) (a) Intensity {\it versus} neutron energy loss at various temperatures after extracting the signals from the raw data. (b) The imaginary part of the dynamical susceptibility (see text) at base temperature at $H=0$ and $3$ T. The solid line is a fit to the data at $H=0$ using sum of six gaussians the positions of which are indicated by arrows. (c)   Illustration of the energy levels of the \{Fe$_6$\} cluster at $T = 0$ K and at higher temperatures, showing the emergence of the peak at 0.5 meV from the partially populated $S^1=1$ to the next $S^2=1$ level.}
\label{fig:temp}
\end{figure}

Figure~\ref{fig:Data-Reduction} shows intensities integrated over the $Q=0.7$ to 1.8 \AA$^{-1}$ range versus neutron energy loss (designated positive) at the lowest temperature achievable with the cryo-magnet, nominally $T=0.1$ K at 0 and 2 Tesla.  To obtain the genuine magnetic spectra from the sample we reduce the data by using a procedure we employed in similar INS studies of the \{Mo$_{72}$Fe$_{30}$\} and \{Cr$_8$\} magnetic molecules.\cite{Garlea2006,Vaknin2010}  The quasi-elastic term representing the instrumental resolution function, the incoherent and static disorder due to the sample, is modeled by a sum of two co-centered peaks: a dominant Gaussian and a minor Lorentzian.  To obtain reliable parameters of the Gaussian/Lorentzian, the spectra at various temperatures are refined simultaneously, with extra peaks due to  magnetic excitations  from the sample, while maintaining the same values of these parameters for the refinements of the various data sets at all measured temperatures, and fields.    By subtracting the co-centered Gaussian/Lorentzian function from data, such as those  shown in the inset of Fig.\  \ref{fig:Data-Reduction}(a) we obtain the intrinsic magnetic excitations from the sample as shown in Fig.\  \ref{fig:Data-Reduction}(a) at 0.1 K at 0 and at 2 T\cite{comment1}.  Using this procedure we obtain the spectra as a function of temperature at zero magnetic field (Fig.\ \ref{fig:temp}) and as a function of applied magnetic field at base temperature Fig.\ \ref{fig:Data-Reduction}(b).  The spectra at base temperature consist of two types of excitations; one that appears at zero magnetic field and the other emerges only with the application of external magnetic field.  We note that the spectra in Fig.\ \ref{fig:Data-Reduction} shows a weak peak at $\sim 0.2$ meV that did not show any magnetic field or temperature dependence. We therefore attribute this excitation to a low energy vibrational and  or rotational excitation of the molecule and therefore is irrelevant to the present study.

Figure\ \ref{fig:temp}(a) shows intrinsic spectra of two excitations centered at  $E_1 \approx 1.0$ and at $E_2 \approx 1.5$ meV at various temperatures that we attribute to $S^1=1$ and $S^2=1$ spin states, as justified below.   Each of the two excitations is  adequately described by three gaussians with peak positions that are evenly displaced at $E_i$ and at $E_i \pm \Delta E_i$  where $\Delta E_1 \approx 0.1$ and $\Delta E_2 \approx 0.12$ meV.  This is demonstrated by a  fit to the data with a sum of six gaussians (solid line) to the imaginary part of the dynamical susceptibility ${\rm Im}\chi(E)$ in Fig.\ \ref{fig:temp}(b) at 0.1 K at zero field.  In our case, the dynamic susceptibility is practically $q$-independent given by\cite{Furrer2013} 
\begin{equation}
{\rm Im}\chi(E) \propto I(E)\left[1-e^{\frac{-E}{k_BT}}\right]^{-1}
\end{equation} 
where $k_B$ is the Boltzmann constant. As the temperature increases, an excitation emerges at exactly the difference between the centers of the two excited states at $E_1-E_2 \approx 0.5$ meV, characteristic of discrete thermally induced excitation from the lower, now populated level  $S^1$, to the upper near-by $S^2$ state.  This new excitation ($E_1-E_2 \approx 0.5$) is narrower than the combined triplet as expected by selection rules between these two $S=1$ states as schematically shown in Fig.\ \ref{fig:temp}(c), namely, $\Delta S =0$ and for the magnetic quantum numbers, $\Delta M = M^1-M^2=0, \pm 1$.   Further evidence that these excitations are magnetic in origin is demonstrated by the strong effect  the magnetic field has on the spectra, in particular in shifting the spectra to lower and higher energies (albeit smeared and impossible to resolve) from the $E_i$s as expected from Zeeman effect on the magnetic quantum states $M$. These observations are consistent with the low-lying magnetic states of the \{Fe$_6$\} cluster as shown schematically in Fig.\ \ref{fig:temp}(c).  While two degenerate excited $S=1$ states  at approximately 1 meV  (with six degenerate magnetic quantum numbers) have been predicted in a previous study  for this molecule\cite{Tolis2006}, the INS presented here shows a more detailed magnetic structure with two $S=1$ states that are split into  their quantum magnetic numbers.  Similar splitting in the spin-excitations of the singlet ground state of SrCu$_2$(BO$_3$)$_2$ has been attributed to the Dzyaloshinski-Moriya  type interaction\cite{Gaulin2004}. Such splitting can also be due to finite coupling between the two \{Fe$_6$\}-\{Fe$_3$\} fragments or due to single-ion anisotropy albeit very weak at that as the $3d^5$ configuration of Fe$^{3+}$ has orbital angular momentum approaching $L = 0$, and hence the second order spin-orbit coupling, that commonly induces a single-ion-anisotropy term, is practically negligible, but cannot be excluded a priori.  Alternatively, magnetic frustration that originates from the triangular building blocks of the \{Fe$_6$\} fragment  (see Fig.\ \ref{fig:molecule}) although as conjectured settle into a total spin $S=0$ for this cluster,\cite{Kahn1997} may yet possess more nuanced excited states as those reported here.  

\begin{figure}
\includegraphics[width=2.1 in]{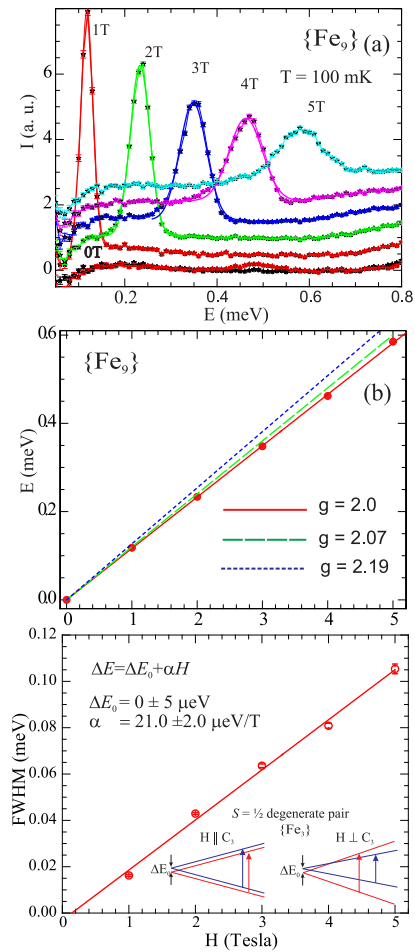}
\caption{(Color online) (a) Intensity versus energy loss at various magnetic fields showing a single excitation that shifty linearly and broadens with the increase in magnetic field. (b) Peak position of the excitations as a function of magnetic field with a linear fit that corresponds to a powder-averaged $g=2.00$. (c) Linewidth of the excitations shown in (a) as a function of magnetic field.  The solid line is a linear fit with parameters included in the figure.}
\label{fig:field2}
\end{figure}
The pronounced magnetic field dependent excitation shown in Fig.\ \ref{fig:field2}(a) indicates effectively a second ground state for the system that naturally is attributed to the \{Fe$_3$ \} fragment of the molecule, suggesting the absence or a very weak coupling between the two fragments.  The excitation shifts to higher energies and broadens with the increase in magnetic field, with a peak position that is perfectly linear in magnetic field with a slope that yields a Zeeman splitting of an $S=1/2$ ground state with a $g = 2.00(2)$ (see Fig.\ \ref{fig:field2}(b)). This is consistent with the theoretical ground state of the tri-nuclear \{Fe$_3$\} fragment with a single exchange coupling among nearest neighbors, namely a doubly degenerate $S=1/2$ ground state\cite{Adams1966} with small variations from the values obtained by EPR measurements\cite{Tolis2006}.  The close to perfect $g= 2.0$ value is a remarkable result as it is an angular average over any crystalline anisotropy for polycrystalline sample with random crystallographic orientation with respect to the applied magnetic field. A possible explanation to the field dependent linewidth shown in Fig.\ \ref{fig:field2}(c) may be explained by assuming a special $g-$value anisotropy reported in Ref.\ \onlinecite{Tolis2006}.  This study reported an effective  $g$-value that depends on the angle between the external magnetic field and the non-crystallographic $C_3$ axis of the 3-spin triangle  such that $g_{\perp} = 2.003$ for $H $ perpendicular, and a pair of values $g_{||} =1.79$ and 2.25 for $H$ parallel to the $C_3$ axis. For a crystalline with angle $\theta$ between the $C_3$ axis and a field direction we write an effective $g$-value   $g_{\rm eff}=2.00 \pm \Delta g\cos(\theta)$, where $\Delta g\approx 0.21$ so that the Zeeman splitting for a single crystal is 
\begin{equation}
E_{S=1/2} = [ 2.00\pm \Delta g\cos(\theta)]\mu_BH.
\end{equation}
Averaging over $\theta$ yields a single value at $\langle E_{S=1/2} \rangle = 2.0\mu_BH$ as observed, with a second moment average, i.e., line-width $\langle \Delta E \rangle \propto \Delta g\mu_BH$, linear in $H$ as shown Fig.\ \ref{fig:field2}(b).  The linear dependence of the FWHM as a function of $H$ shown in Fig.\ \ref{fig:field2}(c) indicates no (or unresolved) energy gap close to the ground state that for the perfect \{Fe$_3$\} triangle consists of a degenerate pair of $S = 1/2$ doublets\cite{Adams1966}.   This behavior suggests that the $S=1/2$ ground state is doubly degenerate as expected from a magnetically frustrated triangle\cite{Adams1966,Kahn1997} diminishing the possibility of a mutual coupling between the two fragments. Based on the linewidth behavior of the excitations at low magnetic fields, we estimate that any splitting due to the removal of this degeneracy is on the order or less than 5 $\mu$eV.   The first excited state for the perfect $s=5/2$ \{Fe$_3$\}  with a single AFM exchange coupling $J$ in a Heisenberg Hamiltonian, (${\cal H} = -J\Sigma {\bf S}_i\cdot {\bf S}_j$) is a total spin $S=3/2$ at $E=3/2J$ from the ground state, and based on the value reported for $J \approx 22$ meV\cite{Tolis2006} is beyond the energy range available with OSIRIS.

To summarize, high resolution inelastic neutron scattering experiments on the\{Fe$_9$\} tridiminshed iron icosahedra magnetic molecule yield comprehensive magnetic spectra that provide the necessary benchmark for  theoretical development of a complete Hamiltonian for this system.  The INS unequivocally shows two sets of ground states, $S=0$ and $S=1/2$, that are attributed to the two \{Fe$_6$\} and \{Fe$_3$\} fragments of the molecule that may be very weakly coupled.  We find that the degeneracy predicted for the first $S=1$ excited states of the \{Fe$_6$\} fragment is removed presumably by finite coupling between the fragments or by inherent frustration in the triangular building blocks of the fragment.   Our magnetic field dependent results at base temperature reveal a remarkable single $S=1/2$ ground state with  $g=2.0$ as expected from a perfect equilateral triangle  \{Fe$_3$\}  fragment. The absence of a clear near by excited state for the  \{Fe$_3$\}  fragment suggests the $S=1/2$ is doubly degenerate as expected from a frustrated triangular arrangement of half-integer spins\cite{Kahn1997,Baker2012}.  More generally, the two fragments  \{Fe$_3$\}  and  \{Fe$_6$\}  with triangular building-blocks may shed light on the effect of topological magnetic frustration in magnetic molecules. 

{\bf Acknowledgments:} {DV thanks Marshall Luban and Larry Engelhardt  for helpful discussions; samples were prepared by Dr Grigore A. Timco (University of Manchester). The work at the Ames Laboratory was supported by the Office of Basic Energy Sciences, U.S. Department of Energy under Contract No.  DE-AC02-07CH11358. We gratefully acknowledge the support by the ISIS cryogenics team.}


\begin{thebibliography}{99}
\bibitem{comment1} Reducing of the genuine signals by deconvoluting the spectrometer's resolution yields practically the same results 
as all peaks are broader than the resolution.
\end{thebibliography}
\end{document}